# Tuning of superconducting properties with disorder in $Nb_xSn$ nanocrystalline thin films

Mahesh Poojary[1], Vishwanadh Bathula[2], Yash Kumar[1], Amar Verma[1], Ekta Kadam[1] and Sangita Bose[1,*]

[1]School of Physical Sciences, UM-DAE Center for Excellence in Basic Sciences, University of Mumbai, Kalina Campus, Santacruz E, Mumbai 400098, India

[2]Materials Science Division, Bhabha Atomic Research Centre, Mumbai 400085, India.

Email: sangita@cbs.ac.in

**Abstract**

Nanocrystalline superconducting films offer an excellent platform to explore the interplay between disorder, granularity, and dimensionality. In this work, we investigate two series of $Nb_xSn$ thin films with near-stoichiometric (x = 3) and slightly Sn-rich (x = 2.5) compositions, deposited on Si (100) substrates via DC magnetron sputtering. Both series exhibit nanocrystalline morphology, with the Sn-rich films displaying smaller grain sizes and a more granular microstructure. A suppression of the superconducting transition temperature ($T_c$) with decreasing film thickness is observed in both series. Notably, a disorder-driven crossover to an insulating state emerges, occurring at a thickness of approximately 11 nm for the Sn-rich films – about twice that of the stoichiometric films. The estimated disorder parameter ($k_F l \approx 0.4$) in the thinnest films indicates proximity to the Anderson localization regime for these films. Magnetotransport measurements reveal a thickness-driven 3D-to-2D crossover, with its onset strongly dependent on film stoichiometry. Furthermore, a pronounced suppression of superfluid stiffness is observed in the Sn-rich films, corroborating the structure-property correlations identified in this study. These work highlights the decisive role of stoichiometry-controlled disorder in tuning superconductivity in granular $Nb_xSn$ thin films.

**Introduction**

Although nanocrystalline or granular superconducting films (GSFs) have been studied for several decades[1,2,3,4,5,6,7] they have recently garnered renewed interest[8,9,10,11] due to their potential to exhibit novel quantum phenomena and their applicability in quantum circuits[12,13,14,15,16,17]. These systems are characterized by the presence of small superconducting grains or islands separated by insulating or weakly conducting matrix, where the intergranular regions act as tunneling barriers and effectively form networks of Josephson junctions[6]. The superconducting state in such systems is therefore governed not only by the intrinsic properties of individual grains but also by the coupling between them. Tuning the grain size and intergrain coupling has been shown to modulate superconductivity, leading to the observation of a superconducting dome, as seen in granular aluminum films[18,19,20]. Multiple reentrant resistive states can also occur in GSFs, influenced by the size and distribution of the superconducting grains[21,22]. These phenomena have been attributed either to electron–electron interactions or to weak localization[12,23]. Furthermore, in systems with relatively small grains and moderately weak intergranular coupling, superconducting fluctuations (SFs) can also play a significant role. In addition, the reduced grain size can lead to increased disorder, which significantly impacts the superconducting properties. In disordered films, the superconducting transition temperature ($T_c$) decreases with increasing disorder and at strong enough disorder, superconductivity can be suppressed altogether. Moreover, disorder has been shown to drive quantum phase transitions (QPTs)[24,25], such as the superconductor-insulator transition (SIT) or the superconductor-metal transition[26], particularly in ultrathin films. These transitions, observed in a variety of systems, are often categorized based on their underlying mechanisms, which may be either "Fermionic" or "Bosonic" in origin[27,28]. Thus, the superconducting properties of GSFs are often governed by a complex interplay of granularity, disorder, and low dimensionality.

$Nb_3Sn$ is a type II superconductor with a high critical temperature ($T_c$) of 18 K and an upper critical magnetic field ($H_{c2}$) of 30 T[29]. In addition to its established use in high-field magnets, $Nb_3Sn$ has also found applications in superconducting radiofrequency (SRF) cavities[30], where thin films grown on copper or niobium substrates are employed to fabricate cavities with high quality (Q) factors[31]. As a well-studied system[32], $Nb_3Sn$ thin films have been extensively investigated for how superconducting properties vary with film thickness[33], particularly focusing on technologically relevant parameters such as $H_{c2}$ and the critical current density ($J_c$)[34,35]. It has been shown that superconductivity is often influenced by composition, morphology and strain in the films.[36,37] However, the influence of stoichiometry-controlled disorder on the evolution of superconducting properties in nanocrystalline or granular $Nb_3Sn$ thin films remain largely unexplored. Whether disorder can drive a SIT/SMT, or induce a dimensional crossover from three-dimensional (3D) to two-dimensional (2D) behaviour in granular $Nb_3Sn$ films, remains an open and intriguing question for this otherwise conventional superconducting system.

In this work, we investigate two series of $Nb_xSn$ thin films with distinct Nb:Sn compositional ratios — one near-stoichiometric (x = 3) and another slightly Sn-rich (x = 2.5) grown on Si (100) substrates by DC magnetron sputtering. In the first series (x = 3), superconductivity persists down to a film thickness of 6 nm below which a transition to a non-superconducting state is observed. In the second series (x = 2.5), the crossover to an insulating state occurs at a higher thickness (~11 nm), highlighting the role of enhanced disorder due to excess Sn. Comprehensive structural and morphological characterizations reveal the nanocrystalline nature of the films, with grain sizes ranging between 10 - 40 nm across both series. However, for the Sn-rich films, the intergranular regions are notably wider in the thinnest samples, indicating reduced inter-grain coupling. Additionally, magnetoresistance measurements and H–T phase diagrams demonstrate a disorder-dependent 3D-2D crossover. A strong suppression of the superfluid stiffness ($J_s$) from electrodynamic measurements is obtained for films with thickness as high as 23 nm for series 2, thereby confirming the role of increased disorder in influencing the superfluid density and affecting the superconducting properties in these films. These results highlight the combined roles of composition, thickness and disorder in governing superconductivity in $Nb_xSn$ thin films. While reduced dimensionality enhances quantum fluctuations in ultrathin films, deviation from stoichiometry introduce additional disorder that can suppress superconductivity at comparatively larger thickness. Our comparative study provides new insight into how compositional disorder controls the critical thickness and superconducting robustness in nanocrystalline $Nb_xSn$ films.

**Experimental Details**

Two series of $Nb_xSn$ thin films (one with x = 3 and another with x = 2.5) were grown using DC magnetron sputtering from a single $Nb_3Sn$ target. The sputter deposition was performed in a custom-built chamber (developed by Excel Instruments). High-purity (99.9%) $Nb_3Sn$ targets, commercially sourced from Testbourne Ltd., were used for the deposition process. The choice of substrate (Si (100), MgO (100), substrate temperature (~800$^0$C) and deposition power (120 – 190 W) were first optimized for the films of the first series (Details given in the Film Growth and Characterization section). Prior to deposition, the chamber was evacuated to a base pressure of ~$2 \times 10^{-6}$ mbar and the deposition of the thin films was performed at an Ar gas pressure of ~$4 \times 10^{-3}$ mbar. The flow rate of the Ar gas was fixed at 30 SCCM for all depositions. For the first series, films with thickness between 1000 – 5 nm was grown and for the second series films of thickness between 100 – 11 nm were grown. The film thickness was measured using a KLA Tencor stylus profilometer on the $Nb_xSn$ thin films deposited in a strip line using a shadow mask.

The structural characterization of $Nb_xSn$ thin films was conducted using a Rigaku Miniflex 600 X-ray diffractometer. The grain size (D) was calculated as the coherently diffracting domain size from XRD peak broadening using the Debye-Scherrer formula after correcting for instrumental broadening. The surface morphology and elemental composition were analysed using scanning electron microscope (SEM) and energy dispersive X-ray analysis (EDX) using a field emission scanning electron microscope (Zeiss ultra FESEM Germany). Transmission electron microscopy (TEM) was conducted using a Cs-corrected FEI Titan microscope, which is operated at 300 kV and equipped with a field emission gun source. To prepare the TEM sample, the $Nb_3Sn$ films were grown onto a TEM grid featuring an array of nine evenly spaced 0.01 $mm^2$ square windows coated with silicon nitride.

The superconducting transition temperature ($T_c$) of the films was measured using both dc transport and the electrodynamic (or the dynamic shielding response) measurements. The dc transport measurements were performed using standard four-probe method with a Keithley 2400 sourcemeter. Hall measurements were carried on the $Nb_xSn$ thin films deposited in a six-probe hall bar geometry using a shadow mask. All these measurements were carried out in a Cryogenics, UK cryogen-free system, capable of reaching temperatures of 300 mK and magnetic fields up to 9 T. The diamagnetic shielding response and the penetration depth ($\lambda$) measurements were done using a custom-designed two-coil mutual inductance probe integrated with the cryogen-free system. The details of the set-up and the measurements are the same as in Ref [38]. The real (imaginary) mutual inductance M′ (″) was measured as a function of temperature from which both $\lambda$ and $J_s$ were obtained as explained in Ref [39].

**Film Growth and Characterization**

$Nb_3Sn$ thin films have been previously synthesized using various techniques including molecular beam epitaxy[40], sputtering[41], evaporation[33], electrochemical synthesis[42]. Among these, magnetron sputtering has been employed using either a single stoichiometric target[43] or separate Nb and Sn targets for co-sputtering or multilayer deposition[44,45]. Most methods have produced high-quality superconducting $Nb_3Sn$ films with critical temperatures ($T_c$) ranging from 15 to 18 K, depending on substrate material and deposition conditions.

In this study, we employed DC magnetron sputtering using a single target to grow $Nb_xSn$ films. $Nb_3Sn$ films (series 1) were deposited on various substrates, including polycrystalline Nb, MgO(100), and Si(100). For films grown at the same sputtering power (122 W) and thickness (100 nm), the film deposited on Si(100) exhibited a higher $T_c$ than that on MgO(100) (see Fig. 1(a), bottom panel). Based on this observation, Si(100) was chosen as the substrate for all subsequent experiments. A high $T_c$ of 18 K was achieved for a 1000 nm thick film grown on a polycrystalline Nb substrate. This demonstrates the availability of the technique to grow high $T_c$ films on Nb substrates used for SRF-based applications. Consistent with previous studies, an increase in substrate temperature was found to enhance the superconducting transition temperature, with a $T_c$ of 17.7 K observed at 800°C[44].

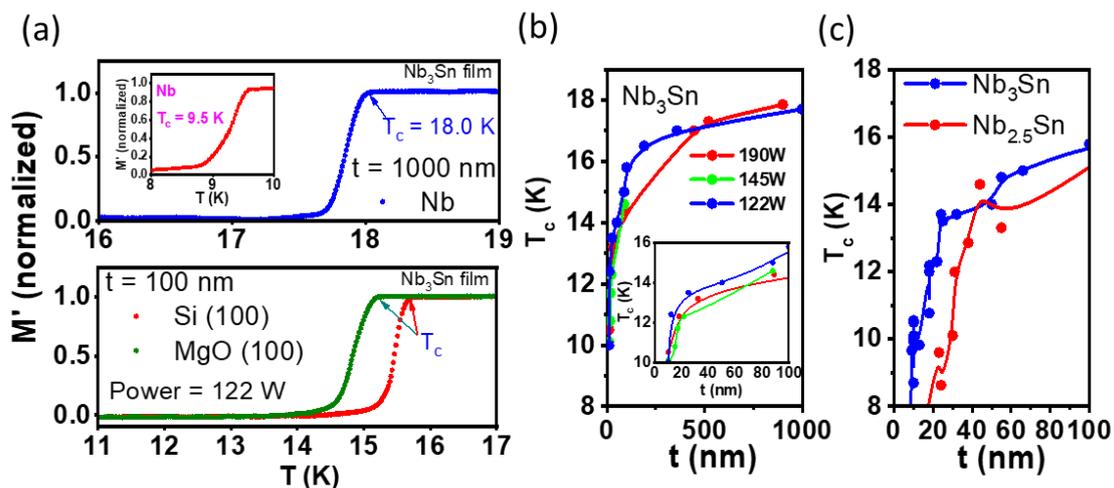

*Figure 1: (a) Real part of normalized mutual inductance (M') vs temperature of $Nb_3Sn$ thin films (a) Top Panel - of thickness, t = 1000 nm on Nb substrate. The inset shows the Nb transition at 9.5 K associated with the Nb substrate, Bottom Panel - of thickness t = 100 nm on MgO(100) and Si(100) substrates. (b) $T_c$ vs t for $Nb_3Sn$ thin films at three different sputtering powers. The inset shows the expanded view of the 0–100 nm thickness region. (c) $T_c$ vs t for $Nb_3Sn$ and $Nb_{2.5}Sn$ thin films grown at 122 W on Si(100) substrate and at a temperature of 800 °C.*

To investigate the influence of film thickness (t) and sputtering power on superconducting properties, $Nb_3Sn$ films (Series 1) were grown at three power levels—122 W, 145 W and 190 W—across a range of thicknesses. For films thinner than 100 nm, those grown at 122 W consistently showed slightly higher $T_c$ (see inset of Fig. 1(b)). In all cases, $T_c$ decreased monotonically with decreasing film thickness, exhibiting a sharp drop below 20 nm. While series 2 films were grown using a Sn-rich $Nb_xSn$ target, resulting in $Nb_{2.5}Sn$ composition, as confirmed by EDS. $Nb_{2.5}Sn$ films (Series 2) deposited on Si(100) substrates (substrate temperature = 800 °C and sputtering power = 122 W) also exhibited a similar variation of $T_c$ with film thickness (Fig. 1(c)).

The XRD patterns of the as grown thin films showed sharp peaks corresponding to the (200), (210) and (211) planes of the $Nb_3Sn$ A15 phase (Figure 2(a)-(b)) indicating the formation of polycrystalline films. Films with lower thickness exhibited broadened peaks, suggesting a reduction in particle size with decreasing film thickness. The grain size was calculated using the Debye-Scherrer formula using the (210) peak, revealing that the grain size varied for different thickness between 40 - 25 nm for series 1 and 36 – 9 nm for series 2 films further confirming the formation of nano-crystalline $Nb_xSn$. Energy dispersive x-ray spectroscopy (EDS) carried on the $Nb_xSn$ films confirmed their composition, with x = 3 for films of series 1 and x = 2.5 for that of films of series 2. Furthermore, SEM images of the films for both series revealed a granular morphology with good homogeneity (see figure 2 (c) – (d)). However, for films with comparable thickness (~15–18 nm), a distinct inter-granular region was observed in the series-2 films with x = 2.5. Films with lower thickness show a similar microstructure (see fig S1 in supplementary information). However, the grain connectivity weakened and the inter-granular contrast became more prominent with decreasing thickness. TEM performed on the $Nb_3Sn$ thin films with thicknesses of 100 nm and 10 nm (see Fig. S2(a) and (d) in the Supplementary Information) showed that the films remained nanocrystalline down to the lowest thickness. This was further confirmed by HRTEM in the 100 nm film (see fig S2 (b) in the Supplementary Information). Moreover, distinct diffraction rings were seen in the selected area electron diffraction (SAED) patterns corresponding to the $Nb_3Sn$ phase (see Fig. S2 (c) and (e) in the Supplementary Information).

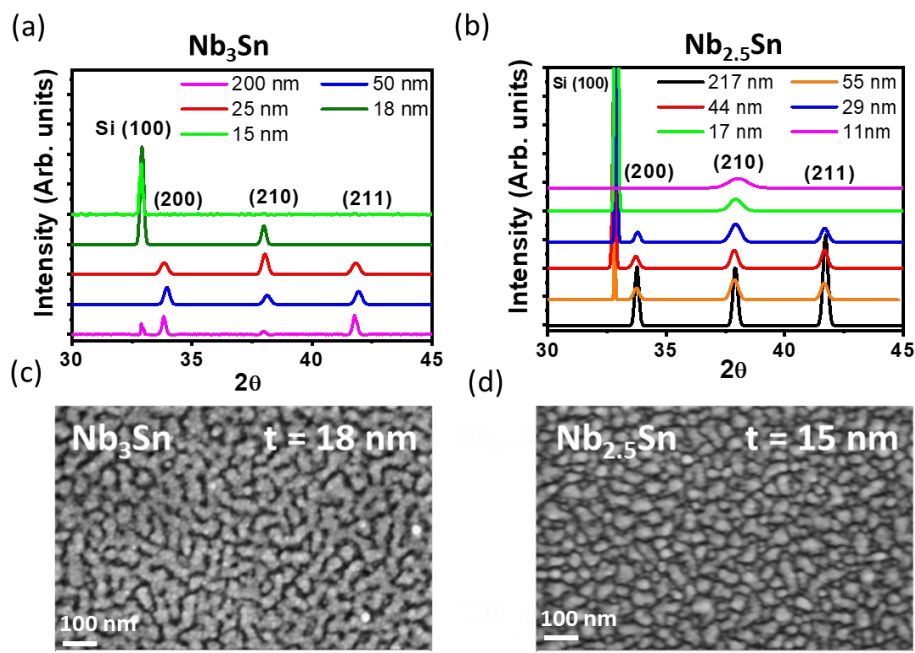

*Figure 2: (a) – (b) XRD pattern of $Nb_xSn$ films with different thickness (t) (x = 3 and x = 2.5 respectively. (c) – (d) SEM images for t = 18 nm and t = 15 nm for $Nb_3Sn$ and $Nb_{2.5}Sn$ films respectively.*

**Result and Discussion**

Figure 3(a)-(b) presents the temperature dependence of the sheet resistance $R_{sq}$ ($T$) at zero magnetic field with varying thickness for films of both series 1 and 2 respectively. In both series, $T_c$ decreases monotonically with reducing thickness, consistent with increasing disorder. Moreover, the superconducting transition remains sharp for thicker films, with a broadening observed at lower thicknesses. In contrast, the off-stoichiometric $Nb_{2.5}Sn$ series (Series 2) shows a more rapid suppression of $T_c$ with decreasing thickness, as seen in Figure 1(c). Interestingly, a disorder-driven crossover to an insulating state is observed for films with reducing film thickness of both series when $R_{sq}$ ($T$) increases beyond the quantum resistance of 6.5 kΩ. Negative temperature coefficient of resistance is observed below 6 nm thick films for Series 1 and below 13 nm for films of series 2. It is worth noting that an increase in sheet resistance in granular films has been shown to indicate reduced inter-grain coupling[6,18,19]. Thus, the observation of the increase in sheet resistance of the $Nb_xSn$ films with decreasing film thickness or different sheet resistance in films of comparable film thickness but different stoichiometry, indicates that the inter-grain coupling changes (This can be corelated with the increase in inter-granular region in thinner films of the $Nb_{2.5}Sn$ films as seen from SEM analysis). Furthermore, this enhanced sensitivity to dimensional tuning along with higher sheet resistance in films of series 2 can be attributed to higher intrinsic disorder. This intrinsic disorder originates from the deviation from the ideal stoichiometry (as confirmed by EDS analysis) giving rise to structural and compositional imperfections like vacancies, interstitials *etc*. The intrinsic effective disorder in the films was quantified by estimating the Ioffe Regel parameter ($k_f l$, where $l$ is the mean free path and $k_f$ is the Fermi wave vector) which was obtained from Hall and transport measurements. In disordered superconductors, it is well established that the dimensionless parameter $k_f l$ decreases with increasing disorder and correlates with the suppression of $T_c$[46,47]. In our $Nb_xSn$ thin films, $k_f l$ spans a wide range, from the moderately disordered ($k_F l \sim 4$) down to $k_F l \sim 0.4$ as seen in Figure 3 (c). Such low values of $k_f l < 1$, are characteristic of systems nearing the Mott-Ioffe-Regel limit, where the semiclassical transport picture breaks down and are widely recognized as indicative of proximity to the Anderson disorder driven metal- insulator transition[48].

To further understand the mechanism governing the variation of $T_c$ with thickness in $Nb_xSn$, we analysed the data using Finkelstein model. In this model, increasing disorder weakens electronic screening, thereby enhancing the diffusive electron-electron coulomb repulsion that counteracts the phonon-mediated pairing interaction and leads to a systematic suppression of $T_c$[49]. Here, $T_c$ is related with the sheet resistance $R_s$ and is given by,

$$\frac{T_c}{T_{c0}} = \exp(\gamma)\left(\frac{1-X}{1+X}\right)^{1/\sqrt{2r}} \quad, \quad r = \frac{e^2}{2\pi^2\hbar}R_s \;, \; X = \frac{\sqrt{r/2}}{\frac{r}{4}+\frac{1}{\gamma}} \;, \; \gamma = ln(\hbar/\tau T_{c0}k_B)$$

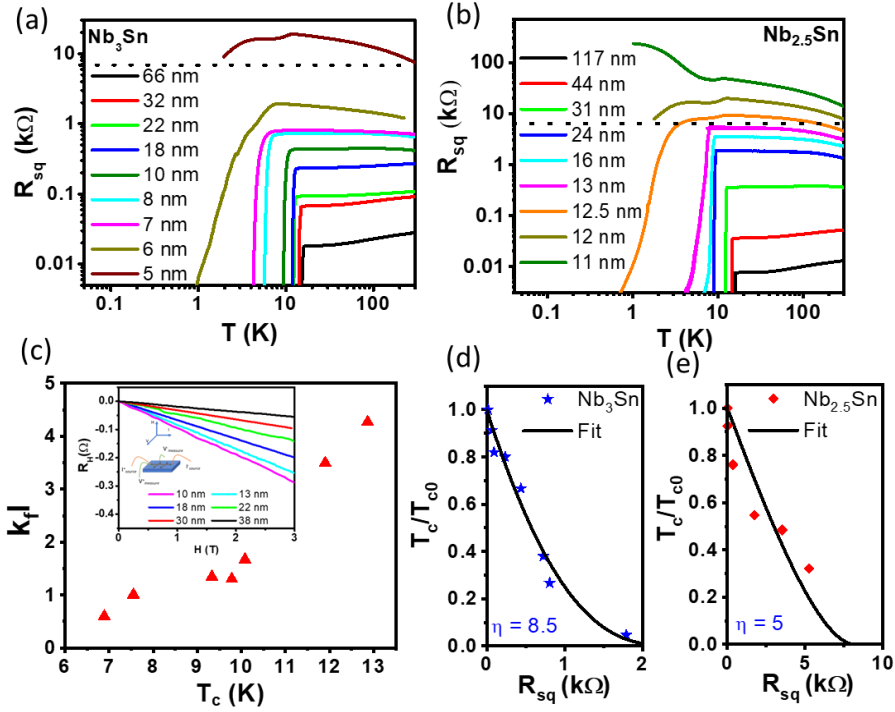

Figure 3: (a) – (b) Variation of sheet resistance ($R_{sq}$) with temperature for $Nb_xSn$ films with different thickness (t) for x = 3 and x = 2.5 respectively. The dotted line in both represents the quantum resistance ($R_q$ = 6.45 kΩ) (c) Variation of $k_Fl$ with $T_c$ for the $Nb_xSn$ films. The inset shows the Hall resistance ($R_H$) with magnetic field (H).
(d) – (e) Variation of normalized $T_c$ with sheet resistance for films of series 1 and 2.

Here $T_{c0}$ is the limiting value of $T_c$ for large thickness. In Figures 3(d)-(e) we show the fit of our data with the model for both the series. For series 1, the fit is reasonably good, with $T_{c0} \sim$ 17 K and $\gamma \approx$ 8.5. In the Finkelstein framework, γ is the fitting parameter characterizes the strength of disorder-enhanced Coulomb repulsion, with larger values indicating stronger depairing due to reduced screening[50,51,52]. The extracted value therefore reflects significant interaction-driven suppression of superconductivity in series 1. In contrast, the poor fit obtained for series 2 suggests that assumptions of homogenous disorder and diffusive transport underlying the Finkelstein model are not fully valid for these films, likely due to additional effects such as granularity, emerging localization or inhomogeneous electronic structure beyond the regime where the theory is applicable.

So far, our observations indicate that the superconducting properties of $Nb_xSn$ films are strongly influenced by their stoichiometry. While $T_c$ decreases with reducing film thickness irrespective of stoichiometry, the disorder-driven crossover to insulating state is found to be stoichiometry dependent. Of particular interest is the 3D - 2D cross-over for films with typical thickness, $d < \xi_0$ (where $\xi_0$ is the coherence length). This crossover can be observed in the temperature dependence of the critical fields. To quantify this behaviour, the temperature variation of the parallel upper critical field (i.e. direction of the applied magnetic field is parallel to the plane of the film, $\mu_0 H_{c2}^{\parallel}$) and perpendicular upper critical field (i.e. direction of the applied magnetic field is perpendicular to the plane of the film, $\mu_0 H_{c2}^{\perp}$) were examined for the $Nb_xSn$ films (see fig. S2 in the Supplementary Information). Fig. 4 (a) −(d) show the phase diagrams for two different thickness for each of the stoichiometry studied. Note: Upper critical fields were defined at 1% (resistance drop with respect to the normal state value) of the respective R (T) curves. An increased anisotropy between $\mu_0 H_{c2}^{\parallel}$ and $\mu_0 H_{c2}^{\perp}$ is observed for decreasing film thickness for both series. While $\mu_0 H_{c2}^{\perp}$ seems to be linear in the temperature range studied for all thickness, the $\mu_0 H_{c2}^{\parallel}$ shows a small curvature. This behaviour of the $\mu_0 H_{c2}^{\parallel}$ (T) curve has been attributed to crossover in thin film systems from 2-dimensions (2D) to 3-

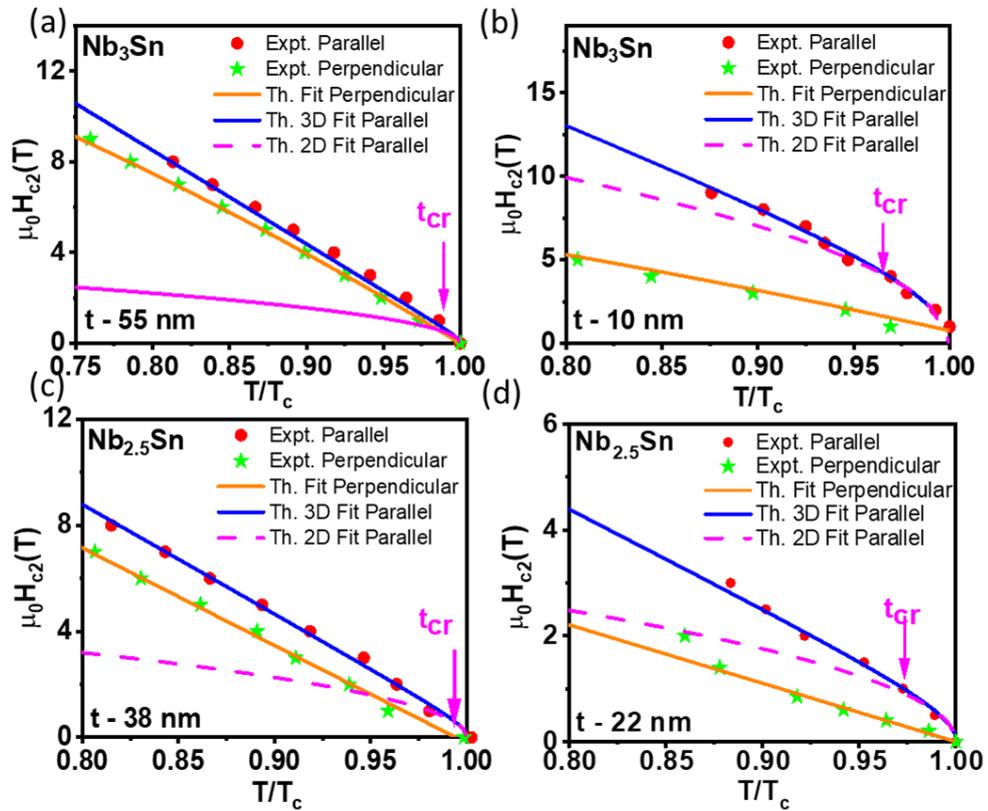

*Figure 4: (a) – (b) Phase Diagrams for two $Nb_3Sn$ films of thickness (t) 55 nm and 10 nm respectively. (c) – (d) Phase Diagrams for two $Nb_{2.5}Sn$ films of thickness (t) 38 nm and 22 nm respectively.*

dimensions (3D) superconductivity[53]. A $T_{cr}$ can be defined which marks the 2D - 3D cross-over temperature. In the 3D regime, d $\gg \xi_0$ for all T $\ll T_{cr}$, while in the 2D regime, anisotropy between the upper critical fields increases at the cross-over temperature, $T_{cr}$. From the GL linearized equations[54,55],

**For 3D regime:**

$$\mu_0 H_{c2}^{\parallel}(T) = \frac{\varphi_0}{2\pi d \xi_0^{\parallel}} \left(\frac{1}{1-\frac{T}{T_c}} + \frac{d^2}{\pi^2 (\xi_0^{\perp})^2}\right)\left(1-\frac{T}{T_c}\right) \quad\quad 1$$

$$\mu_0 H_{c2}^{\perp}(T) = \frac{\varphi_0}{2\pi \xi_0^2}\left(1-\frac{T}{T_c}\right) \quad\quad 2$$

**For 2D regime:**

$$\mu_0 H_{c2}^{\parallel}(T) = (1.05)\frac{\varphi_0}{2d\xi_0^{\parallel}}\left(1-\frac{T}{T_c}\right)^{1/2} \quad\quad 3$$

$$\mu_0 H_{c2}^{\perp}(T) = \frac{\varphi_0}{2\pi (\xi^{\parallel}(T))^2} \quad\quad 4$$

Equations, 2-4 were used to fit the experimental data (Shown in figure 4). For thinner films for both stoichiometry, $t_{cr}$ ($T_{cr}/T_c$) is pushed to lower values. Surprisingly, for off-stochiometric films, this is observed at a much higher film thickness. Thus, the 3D-2D cross-over regime gets affected by the intrinsic disorder in the $Nb_xSn$ films.

To convincingly establish if the disorder is increasing for the films of series 2, superfluid stiffness ($J_s$) was measured using the two-coil mutual inductance measurements[39], (see fig. S3 in the Supplementary Information). Figure 5 shows the temperature variation of the $J_s$ for two different thickness for each of the stoichiometry studied of the $Nb_xSn$ films. As expected for both stoichiometry the superfluid stiffness decreased with the reduction in film thickness for both the series[39].

However, in the $Nb_{2.5}Sn$ (72:28) series, the suppression of $J_s$ is even more pronounced. Both the 55 nm and 23 nm films show substantially reduced superfluid stiffness compared to their series 1 counterparts. This indicates that series 2 possesses an inherently higher disorder level, likely due to stoichiometric off-balance and associated structural inhomogeneity, which reduces the effective superfluid density[39,56,57]. Thus, $J_s$ reflects a combined effect of increased disorder, reduced carrier density and weakened inter-grain coupling in the $Nb_xSn$ nanocrystalline films. It is well established that suppression of $J_s$ makes a superconductor more susceptible to phase fluctuations[58]. Typically, this effect becomes significant when $J_s$ is of the order of or less than the superconducting energy gap. However, the substantial reduction in $J_s$ indicates a loss of global phase rigidity and signifies an enhancement of phase fluctuations in the $Nb_{2.5}Sn$ films at thickness as large as 23 nm.

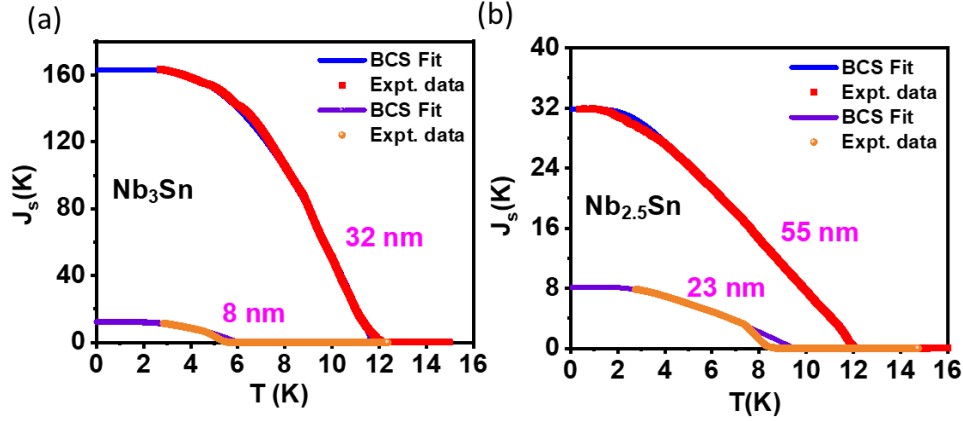

*Figure 5: (a) Temperature dependence of superfluid stiffness ($J_s$) for 32 nm and 8 nm thin films respectively for $Nb_3Sn$ films. (c) Temperature dependence of $J_s$ for 55 nm and 23 nm thin films respectively for $Nb_{2.5}Sn$ films.*

Thus, unlike earlier studies that mainly focused on optimizing $T_c$, $H_{c2}$ or critical current density, the present work demonstrates how stoichiometry-induced disorder strongly modifies superconducting behavior, leading to: (i) a superconductor–insulator transition (SIT) at significantly larger thickness in Sn-rich films, (ii) a disorder-driven 3D–2D dimensional crossover, and (iii) a pronounced suppression of superfluid stiffness in the disordered Sn-rich films. Thus, this work introduces a new material to explore disorder-driven quantum phase transitions that have not been reported previously.

**Conclusion:**

We have systematically investigated the effects of thickness and stoichiometry on the superconducting properties of nanocrystalline $Nb_xSn$ thin films. Both near-stoichiometric (x = 3) and Sn-rich (x = 2.5) films exhibit a monotonic decrease in $T_c$ with decreasing thickness, consistent with enhanced disorder and reduced inter-grain coupling. However, the Sn-rich films show a pronounced suppression of superconductivity, leading insulating behaviour at relatively larger thicknesses. Structural and transport analyses confirm that the increased disorder, quantified by low $k_Fl$ values, is primarily responsible for this behaviour. The observation of a 3D–2D dimensional crossover, supported by anisotropic upper critical field measurements, further illustrates the role of disorder and reduced dimensionality in controlling the superconducting phase. Electrodynamic measurements show a strong reduction in the superfluid stiffness for the Sn-rich series, reinforcing the influence of disorder on the superfluid density. Overall, the contrasting behaviour between the two-film series highlights how compositional deviation from stoichiometry introduces intrinsic disorder that governs the evolution from bulk-like superconductivity to a disordered, low-dimensional regime in Nb$_3$Sn thin films.

**Acknowledgements**

SB acknowledges Department of Science and Technology (DST)- Nanomission (Grant No.: DST/NM/TUE/QM-8/2019(G)/3) and Department of Atomic Energy (DAE), Government of India. All experimental measurements presented here along with the analysis were done by M. P. under the supervision of S. B. Some initial measurements were done by Y. K. Extensive morphological and structural characterization of the films were done by V B. Occasional help in film growth and measurements was obtained from A.V. and E. K. The manuscript was written by S. B. after discussions with all authors. The project was conceptualized by S. B.